\documentclass[journal]{IEEEtran}

\usepackage{cite}

\usepackage{graphicx}

\usepackage{amsmath}

\begin{document}

\title{Demonstration of a passive sub-picostrain fiber strain sensor}

\author{Jong H. Chow,
        Ian C. M. Littler\dag,
        David E. McClelland,
        and~Malcolm~B.~Gray
\thanks{This research was supported by the Australian Research Council (ARC)}%
\thanks{The authors are with the Centre for Gravitational Physics, Faculty of Science,
The Australian National University, Canberra, ACT 0200, Australia;
\dag Ian C.M. Littler is with CUDOS, School of Physics, A28, University of Sydney,
Camperdown, NSW 2006, Australia.}}

\markboth{To appear in Optics Letters}{Shell \MakeLowercase{\textit{Jong H. Chow et al.}}:}

\maketitle


\begin{abstract}
We demonstrate a fiber Fabry-Perot (FFP) sensor capable of detecting sub-picostrain
signals, from 100 Hz and extending beyond 100 kHz, using the Pound-Drever-Hall frequency
locking technique.  A low power diode laser at 1550 nm is locked to a free-space reference
cavity to suppress its free-running frequency noise, thereby stabilizing the laser.  The
stabilized laser is then used to interrogate a FFP where the PDH error signal yields the
instantaneous fiber strain.
\end{abstract}


\IEEEpeerreviewmaketitle


\PARstart{F}{iber} Bragg gratings play an emerging role in the realization of
ultra-sensitive static and dynamic strain detectors for a variety of applications, such as
underwater acoustic array sensors \cite{hill}, embedded monitoring of smart structures in
civil and aerospace industries \cite{kersey, lissak, allsop}, ultrasonic hydrophones for
medical sensing \cite{fisher}, and seismic sensors for geophysical surveys
\cite{SchmidtHattenberger}.  The benefits over the piezo-electric strain sensors currently
employed include their smaller cross-sectional area and their scalability to large arrays.
In addition, the detector arrays could be remotely interrogated and optically multiplexed
using standard telecommunications equipment.

The Pound-Drever-Hall (PDH) frequency locking technique uses RF phase modulation of a laser
beam incident on a Fabry-Perot interferometer.  By detecting and demodulating the beam
reflected off the Fabry-Perot cavity, a high signal-to-noise error signal is derived,
yielding the instantaneous frequency difference between the laser frequency and the cavity
resonance \cite{drever,black}.  This scheme has found many applications in areas involving
laser stabilization and signal extraction \cite{day,deVine}.  The most demanding
application of the PDH technique is in the detection schemes for gravitational waves, which
require strain sensitivities approaching $\Delta L/L = 10^{-22}$ $\varepsilon /
\sqrt{\mathrm{Hz}}$ \cite{abott}, where $\varepsilon$ is a dimensionless unit of strain.

In this paper, the PDH technique is applied to a simple fiber Fabry-Perot interferometer
(FFP) formed by a Bragg grating pair.  We present a fiber sensing system with, to our
knowledge, an unprecedented strain sensitivity of better than $10^{-12}$ $\varepsilon /
\sqrt{\mathrm{Hz}}$, in a band extending down to 100 Hz.

\begin{figure}[htb]
  \centering
  \includegraphics[width=3.0in]{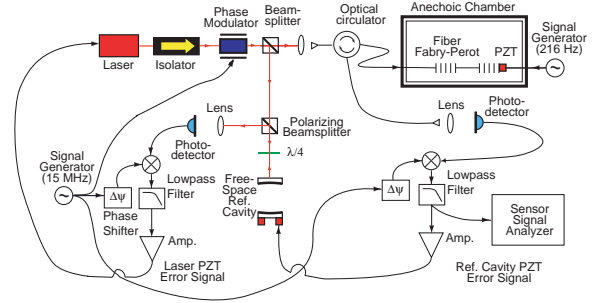}
  \caption{Experimental set-up showing the laser and stabilization cavity together with
  the FFP sensor in the anechoic chamber. The 15 MHz signal generator was used to modulate
  the laser as well as providing the local oscillator for demodulation electronics.
  The 216 Hz signal generator was used to generate the calibration signal for the FFP sensor.}
  \label{schematic}
\end{figure}

The power of the PDH technique lies in its shot noise limited closed-loop spectral density
of frequency noise, which is given by \cite{day}
\begin{eqnarray}
    S_{f, cl min} ~(\mathrm{Hz/\sqrt{Hz}}) = \frac{\Delta\nu_{c}}{J_{0}(\beta)}
                                        \sqrt{\frac{h \nu}{8 \eta P_{i}}},
    \label{eqn_freq_noise}
\end{eqnarray}
where $\Delta\nu_{c}$ is the full-width half-maximum (FWHM) linewidth of the FFP resonance;
$\beta$ is the modulation depth; $\nu$ is the optical frequency of the laser carrier;
$\eta$ is the photodectector quantum efficiency; and $P_{i}$ is the input interrogating
optical power.  Eqn. \ref{eqn_freq_noise} can be readily adapted to provide a theoretical
limit to strain sensitivity in a FFP sensor, such that
\begin{eqnarray}
    \mathrm{Strain} ~(\varepsilon/\sqrt{\mathrm{Hz}}) = \Delta\nu_{c}
                                                            \sqrt{\frac{h}{8 \eta \nu P_{i}}},
    \label{eqn_strain_limit}
\end{eqnarray}
where we have assumed that $\beta$ is small and $J_{0}(\beta)\simeq1$.  Using parameters of
$P_{i} = 1$ mW, $\eta \simeq 0.9$, and $\Delta\nu_{c} = 100$ MHz, with an interrogating
laser wavelength of 1.55 $\mu$m, Eqn. \ref{eqn_strain_limit} yields a shot noise limited
strain sensitivity of $2 \times 10^{-15}$ $\varepsilon/\sqrt{\mathrm{Hz}}$.

Realistic PDH experiments are usually limited to sensitivities far worse than those
predicted by Eqns. \ref{eqn_freq_noise} and \ref{eqn_strain_limit}, due to the free running
frequency noise of the interrogating laser.  In our experiment, therefore, we frequency
stabilized our diode laser to a free-space reference cavity, in order to suppress this
dominant noise source by more than three orders of magnitude.  In addition, our reference
cavity has one mirror bonded to a PZT actuator, allowing us to tune and lock the stabilized
laser to the FFP resonance.  The PDH error signal from the FFP then yields a low noise,
instantaneous fiber strain measurement.  This two-loop control scheme is illustrated in
Fig. \ref{schematic}.  The beamsplitter after the phase modulator tapped off the laser beam
for pre-stabilization, while the main beam was transmitted to interrogate the FFP.

The FFP was mounted on a stage which could be remotely driven to stretch-tune the FFP into
resonance with the interrogating laser. In addition, one end of the FFP was attached to a
PZT, allowing the length to be modulated for a direct calibration of the strain
sensitivity. The FFP and mount were enclosed in a hermetically sealed, mechanically
isolated anechoic chamber to minimize acoustic noise in the FFP.

An external cavity diode laser (New Focus Vortex Model TLB-6029) operating at 1550 nm was
locked, via the PDH technique, to a free-space confocal Fabry-Perot reference cavity with a
free-spectral-range (FSR) of 3 GHz. The cavity was composed of an INVAR spacer with a
bonded PZT actuator to displace one of the end-mirrors, allowing dynamic tuning of the
reference cavity.  The phase modulation for the PDH control loop was applied using a
resonant phase modulator (New focus Model 4003)  operating at 15 MHz. The Fabry-Perot
reference cavity had a FWHM of 35 MHz, yielding an error signal slope of 0.11 $\mu$V/Hz.
The optical power at the laser output was 5 mW and was split between the reference cavity
and the FFP. The diode laser had two frequency actuators: an internal PZT tuning element
with a bandwidth of approximately 3.5 kHz, as well as current feedback with a 1 MHz
bandwidth.

\begin{figure}[htb]
  \centering
  \includegraphics[width=3.0in]{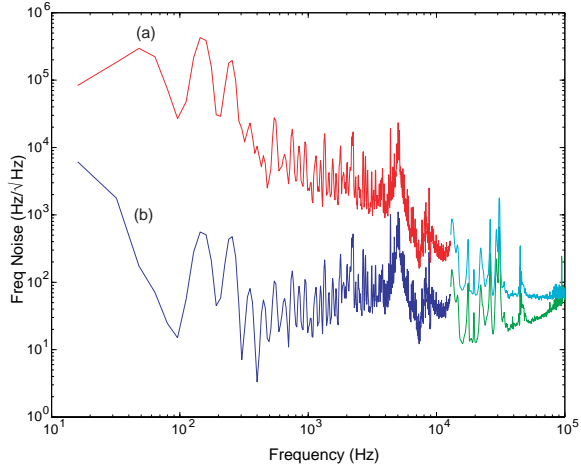}
  \caption{Laser pre-stabilization illustrated by (a) free running and (b) closed-loop laser
  frequency noise, showing noise suppression by up to 3 orders of magnitude in the range 10 Hz
  to 100 kHz.}
  \label{noise_suppression}
\end{figure}

Our laser pre-stabilization servo was designed to have a tailored controller response with
a cross-over from PZT to current feedback at 1 kHz, with unity gain at approximately 100
kHz. This delivered an in-loop laser frequency noise suppression of more than 1000 at 100
Hz, and rolling off to a factor of approximately 10 at 10 kHz. The laser frequency noise
suppression plots are shown in Fig. \ref{noise_suppression}, where the free-running (upper
trace) is overlaid with the stabilized laser noise (lower trace).  The spectral features at
around 2, 5, 9 and 15 kHz were due to mechanical resonances of the laser tuning PZT, while
the features at 50 Hz and associated harmonics, especially the third and fifth harmonics at
150 Hz and 250 Hz respectively, were caused by direct electrical mains pickup.  In addition
there were also features caused by acoustic excitation and mechanical pickup of the
free-space reference cavity.

\begin{figure}[htb]
  \centering
  \includegraphics[width=3.0in]{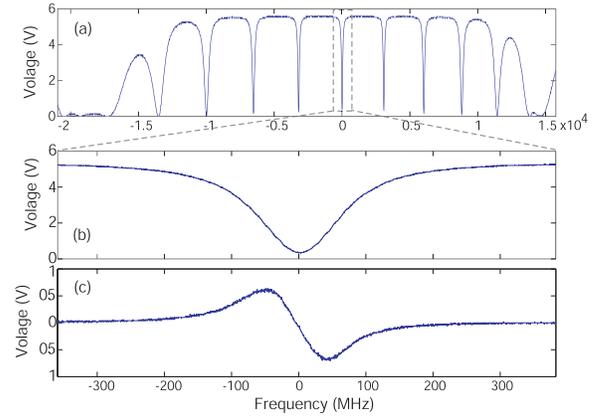}
  \caption{a) The reflection spectrum for the FFP showing 9 supported modes, obtained by
  scanning the laser frequency using its PZT actuator.  The apparent change in FSR is due
  to PZT nonlinarity.  b) Reflection power for the central high finesse FFP mode.  c) its
  corresponding PDH error signal.}
  \label{FFP_scan}
\end{figure}

The FFP was designed to have a FSR of 4.8 GHz, as shown in Fig. \ref{FFP_scan}a, and was
written in hydrogenated Corning SMF-28 fiber using a holographic UV writing technique. It
consisted of two Bragg reflectors, each 3 mm long and separated by 20 mm, with nominal peak
reflectivity of 94$\%$.  The section of fiber between the Bragg mirrors was unexposed to
UV. The Bragg reflectors both had a bandwidth of approximately 40 GHz (320 pm) such that 9
modes could be supported, each with a different finesse due to the change in reflectivity
towards the Bragg band edges.  In this experiment, we used the center high finesse mode as
shown in Fig. \ref{FFP_scan}b, with a FWHM of about 100 MHz, to provide the greatest strain
responsivity.  Its corresponding error signal, derived via PDH demodulation, is shown in
Fig. \ref{FFP_scan}c.  The error signal slope across the resonance was measured to be 20
nV/Hz.  The FFP was close to impedance matched, such that at line center less than $5\%$ of
the incident light was reflected.  The laser power incident on the FFP was typically 1 mW.

To measure the fiber strain sensitivity, the FFP was stretch-tuned such that the central
high finesse mode was nearly resonant with the stabilized laser.  The laser was then locked
to this FFP resonance by feeding back the FFP error signal to the free-space reference
cavity, with a unity gain bandwidth of approximately 20 Hz.

\begin{figure}[htb]
  \centering
  \includegraphics[width=8cm]{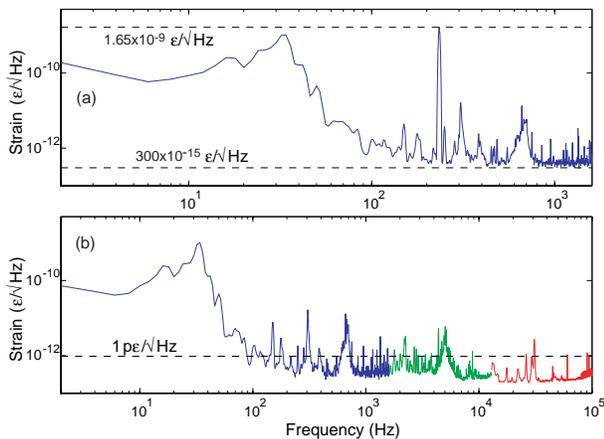}
  \caption{The calibrated noise spectrum of the Fiber Fabry-Perot sensor. In a) an external
mechanical signal of 0.34 nm at 216Hz was applied to the FFP via a PZT.  The background
noise floor is shown in (b), where the noise was due to residual laser noise, fiber violin
modes as well as direct electrical pick-up.  The plots are overlaid with dotted lines for
strain references in the vertical scale.}
  \label{freq_noise}
\end{figure}

To calibrate the strain of this sensor, a signal of 70 mV$_{rms}$ at 216 Hz was applied to
the PZT attached to one end of the FFP, which gave rise to a large peak at 216 Hz, as seen
in Fig. \ref{freq_noise}a.   Knowing that the PZT had a responsivity of 4.8 nm/V at
frequencies below the 600 Hz mechanical resonance, the modulation corresponds to a
displacement of 0.34 nm.  Taking into consideration the 104 mm length of fiber between the
supporting chucks, this displacement equates to an applied strain of $3.3\times10^{-9}$
$\varepsilon$.  After taking the measurement bandwidth of 4 Hz into account, the equivalent
strain spectral density was $1.65\times10^{-9}$ $\varepsilon/\sqrt{\textrm{Hz}}$.  This
signal was then used to calibrate the vertical axes of Fig. \ref{freq_noise}.

From Fig. \ref{freq_noise}a, we see that the signal at 216 Hz was a factor of
$5\times10^{3}$ above the noise floor, yielding a strain sensitivity of $340\times10^{-15}$
$\varepsilon/\sqrt{\textrm{Hz}}$ around 216 Hz.

The broadband noise spectrum of the FFP error signal is shown in Fig \ref{freq_noise}b,
where the calibration signal was turned off.  It shows the existence of multiple features
in the spectrum, due to the laser PZT resonances at around 2, 5, 9 and 15 kHz, fiber chuck
resonance at 600 Hz, fiber violin vibrational modes, as well as electrical pick-up due to
the 50 Hz AC line.  The large wall in the spectrum below 100 Hz was likely due to pick-up
of low frequency laboratory rumble, in the reference cavity used to stabilize the diode
laser. Although the cavity was isolated from high frequency noise with a damping enclosure,
it was still physically attached to the optical table and was thus susceptible to its
mechanical and seismic pollution.  We note that this spectral feature was absent in Fig.
\ref{noise_suppression}, and thus not observed in the in-loop error signal of the laser
itself.  The flat noise floor underlying Fig. \ref{freq_noise}b was present even with the
laser turned off.  Detailed investigations showed that this limiting noise source was due
to electronic noise of the mixer pre-amplifier stage.

The stabilized frequency noise of the laser, seen in Fig. \ref{noise_suppression} trace
(b), produced an equivalent noise floor below $300 \times 10^{-15}$
$\varepsilon/\sqrt{\textrm{Hz}}$, and is therefore not visible in Fig. \ref{freq_noise},
apart from the laser PZT resonances at 2 kHz and 5 kHz.

For convenience in this experiment, we used an interrogation laser power of 1 mW.  This
power level yields a shot noise limited strain sensitivity of $2 \times 10^{-15}$
$\varepsilon / \sqrt{\textrm{Hz}}$, as calculated using Eq. \ref{eqn_strain_limit}, which
is approximately 2 orders of magnitude below the noise floor measured.  Hence we can reduce
the FFP interrogation power down to $\simeq$ 100 nW before the shot noise limited strain
sensitivity becomes comparable to the noise floor of Fig. \ref{freq_noise}.

With the exception of a few well identified noise sources at 300 Hz, which was the fiber
violin mode; 600 Hz which was the fiber chuck resonance; 2 kHz and 5 kHz, which were the
laser PZT resonances; and 30 kHz which was the reference cavity PZT resonance, Fig.
\ref{freq_noise}b demonstrates broadband sub-picostrain sensitivity from 100 Hz to 100 kHz.



\begin{thebibliography}{1}
\bibitem{hill} D. J. Hill, P. J. Nash, D. A. Jackson, D. J. Webb, S. F. O'Neill, I. Bennion,
and L. Zhang, "A fiber laser hydrophone array", Proceedings of SPIE Vol. 3860 (2003).

\bibitem{kersey} Alan D. Kersey, Michael A. Davis, Heather J. Patrick, Michel LeBlanc, K. P.
Koo, C. G. Askins, M. A. Putnam, and E. Joseph Friebele, "Fiber Grating Sensors", J.
Lightwave Technol., \textbf{15}, 8 (1997).

\bibitem{lissak} B. Lissak, A. Arie, and M. Tur, "Highly sensitive dynamic strain measurements
by locking lasers to fiber Bragg gratings", Opt. Lett., \textbf{23}, 24 (1998).

\bibitem{allsop} T. Allsop, K. Sugden, I. Bennion, R. Neal, and A. Malvern, "A high resolution
fiber Bragg grating resonator strain sensing system", Fiber and Integrated Optics,
\textbf{21}, 205-217, (2002).

\bibitem{fisher} N. E. Fisher, D. J. Webb, C. N. Pannell, D. A. Jackson, L. R. Gavrilov,
J. W. Hand, L. Zhang, and I. Bennion, "Ultrasonic hydrophone based on short in-fiber Bragg
gratings", Appl. Opt., \textbf{37}, 34 (1998).

\bibitem{SchmidtHattenberger} Cornelia Schmidt-Hattenberger, Gunter Borm, and F. Amberg,
"Bragg grating seismic monitoring system", Proc. SPIE Vol. 3860, p. 417-424 (2003)

\bibitem{drever} R. W. P. Drever, J. L. Hall, F. V. Kowalski, J. Hough, G. M. Ford,
A. J. Munley, and H. Ward, "Laser phase and frequency stabilization using an optical
resonator", Appl. Phys. B, \textbf{31}, pp. 97-105 (1983).

\bibitem{black} Eric D. Black, "An introduction to Pound-Drver-Hall laser frequency
stabilization", Am. J. Phys., \textbf{69}, pp. 79-87 (2001).

\bibitem{day} Timothy Day, Eric K. Gustafson, and Robert L. Byer, "Sub-Hertz relative
frequency stabilization of two-diode laser-pumped Nd:YAG lasers locked to a Fabry-Perot
interferometer", IEEE J. Quantum Electron., \textbf{28}, pp. 1106-1117 (1992).

\bibitem{deVine} Glenn de Vine, John D. Close, David E. McClelland, and Malcolm B. Gray,
"Pump-probe differencing technique for cavity-enhanced noise-cancelling saturation laser
spectroscopy", Opt. Lett., Accepted, to appear in 2005.

\bibitem{abott} B. Abbott et al, "Detector description and performance for the first
coincidence observations between LIGO and GEO", Nuclear Instr. and Methods in Phys. Res.
\textbf{A517}, 154-179 (2004).

\end{thebibliography}
\end{document}